\documentclass[aps,twocolumn,nofootinbib]{revtex4}
\usepackage{mathrsfs} \usepackage{amsmath} \usepackage{graphicx}
\usepackage{dcolumn} \usepackage{bm} \usepackage{amssymb}
\usepackage{latexsym}

\begin{document}

\title{Does the mass of a black hole decrease due to the
accretion of phantom energy? }

\author{Changjun Gao} \email{gaocj@bao.ac.cn}
\author{Xuelei
Chen}
\email{xuelei@cosmology.bao.ac.cn}
\affiliation{The  National Astronomical Observatories, Chinese
Academy of  Sciences, Beijing, 100012, China}
\author{Valerio Faraoni}
\email{vfaraoni@ubishops.ca}

\affiliation{Physics Department, Bishop's University, 2600
College Street, Sherbrooke, Qu\'{e}bec, Canada J1M~1Z7}

\author{You-Gen Shen}
\email{ygshen@center.shao.ac.cn}
\affiliation{Shanghai Astronomical Observatory, Chinese Academy
of Sciences, Shanghai 200030, China} \affiliation{Joint
Institute for Galaxy and Cosmology of SHAO and
  USTC, Shanghai 200030, China}

\date{\today}

\begin{abstract}
According to Babichev {\em et al.}, the accretion
of a phantom test fluid onto a Schwarzschild black hole will
induce the mass of the black hole to decrease, however the
backreaction was ignored in their calculation. Using new exact
solutions describing black holes in
a background Friedmann-Robertson-Walker universe, we find that
the physical black hole mass may instead {\em increase} due to
the accretion of phantom energy. If this is the case, and
the future universe is dominated
by phantom dark energy, the black hole apparent horizon and the
cosmic apparent horizon will eventually coincide and, after that, the black
hole singularity will become naked in finite comoving time
before the Big Rip occurs, violating the Cosmic Censorship
Conjecture.
\end{abstract}

\maketitle

\section{Introduction}

The first year WMAP data combined with the \textrm{2dF} galaxy
survey and the supernova Ia data favor the phantom energy
equation of state of the cosmic fluid $w\equiv P/\rho <-1$
(where $\rho$ and $P$ are the cosmic energy density and
pressure, respectively) over the Einstein cosmological constant
($w=-1$)  and the quintessence field ($w>-1$)
\cite{ps:04,alam:04}. The interpretation of this result brings
about the possibility that dark energy could be in the form of a
phantom field \cite{caldwell:02,gibbons:03,nojiri:03}. Compared
to the Lagrangian density of a quintessence field $
\mathscr{L}_{quintessence} =
\frac{1}{2}\partial_\mu\phi\partial^{\mu}\phi +
V\left(\phi\right)$, the Lagrangian density of the phantom field
has a kinetic energy term with the ``wrong'' sign,
{\em i.e.}, $
\mathscr{L}_{phantom} =
-\frac{1}{2}\partial_\mu\phi\partial^{\mu}\phi
+V\left(\phi\right) $. This``wrong'' sign endows the phantom
model of dark energy with two surprising properties. First, the
density of phantom matter increases with the expansion of the
universe in such a way that both the scale factor and the
phantom energy density can become infinite in a finite comoving
time. If this happens, every object in the universe, including
galaxies, stars, the earth and nuclei will be torn apart in a
finite future in a so-called Big Rip singularity
\cite{caldwell:02,caldwell:031,nes:04}. A second surprising
property of the phantom field was found by Babichev, Dokuchaev
and Eroshenko \cite{babi:04}, who showed that accretion of
phantom energy will induce the mass of black holes to decrease.
They concluded that the masses of all black holes will vanish
before the Big Rip is reached. However, in their calculations,
they ignored the backreaction of phantom matter on the black
hole metric. In a low matter density background, this effect can
be safely ignored, but if the matter density of the background
is large, for example if it becomes comparable to the black hole
density $ \sim 1/M_0^2$ ($M_0$ is the black hole mass), the
metric describing this black hole will be modified
significantly; hence, the backreaction must be accounted for in
this case.

The purpose of the present paper is to point out that, taking
into account the backreaction, the physical black hole mass may
not decrease but will rather {\em increase} due to
the accretion of
phantom matter. The size of the black hole apparent horizon (AH)
increases, while that of the cosmic AH decreases, during the
evolution of the  universe to a Big Rip. There exists
a moment
$t_*$ at which the black hole AH coincides with the cosmological
AH, and after which the black hole singularity becomes naked. A
naked singularity is forbidden by the Cosmic Censorship
Conjecture \cite{pen:69}; in a broad sense, if the
latter is correct, it seems that the Cosmic Censor
disallows the existence in nature of not
only naked singularities, but also of the cryptic
phantom field.

In order to study this issue we use new exact solutions of the
Einstein field equations describing dynamical black holes
embedded in a Friedmann-Robertson-Walker (FRW)
universe driven by phantom energy and
accreting this cosmic fluid.  Independent motivation for the
development of dynamical black hole solutions comes from various
areas of theoretical physics including black hole
thermodynamics, quantum gravity, and mathematical physics
\cite{ward:2000,bak:99, cai:05, recent, DiCriscienzoetal07,
mybh}. Similar issues about the accretion of phantom energy onto
wormholes were discussed in \cite{wormholes}.

We use units in which $c=G=\hbar=1$ and this paper is organized
as
follows: in Sec.~II we discuss how the McVittie
solution can not
describe a black hole embedded in a FRW background universe.
Then, we turn to new solutions and show that a new metric found
(but not studied in detail) in \cite{fa:2007} can describe a
dynamical black hole in a universe driven by dark matter,
radiation, or dark energy. In Sec.~III, using the solution of
\cite{fa:2007}, we derive the result that the black hole mass
will {\em increase} due to the accretion of phantom matter.
Secs.~IV and VI find the same result for
qualitatively different classes of exact
solutions, while Sec.~VI contains a discussion and
the conclusions.

\section{New cosmological black hole solutions}

To investigate the relation between cosmic expansion and local
physics, McVittie introduced in 1933 his well-known solution
 \cite{mc:1933}
\begin{eqnarray}
ds^2&=&-\frac{\left[1-\frac{M_0}{2a\left(t\right)r}\right]^2}{
\left[1+\frac{M_0}{2a\left(t\right)r}\right]^2} \, dt^2
  +{a^2\left(t\right)}\left[1+\frac{M_0}{2a\left(t\right)r}
  \right]^4
   \nonumber\\ &&\nonumber \\
  &&\cdot\left(dr^2+r^2d\Omega^2 \right) \;,
 \end{eqnarray} where $d\Omega^2$ is the line element on the
unit 2-sphere and the constant $M_0$ denotes the physical mass
of the black hole when $a(t)=$const. (this can be seen by
rescaling the radial coordinate as $\tilde{r}\equiv a{r}$). The
scale factor $a(t)$ is an arbitrary (positive) function of the
comoving time $t$. This metric reduces to the familiar spatially
flat FRW metric as $M\rightarrow
0$.
So at a first glance, the metric seems to represent a
Schwarzschild black hole embedded in a spatially flat FRW
universe \cite{noe:1971}. However it is argued that, in general,
the McVittie solution can not describe a black hole because it
is singular on the 2-sphere $r=M_0/2$ (which reduces to the
Schwarzschild horizon if $a=$const.) \cite{sus:1985} and this
singularity is spacelike \cite{no:1999}. It is also claimed that
the McVittie metric may describe a point mass located at $r =0$
and embedded in a FRW universe. However, this point mass is, in
general, surrounded by the singularity at $r = M_0/2$, which is
difficult to interpret. This singularity was studied in
Refs.~\cite{sus:1985}. Nolan \cite{no:1998} showed that this is
a weak singularity in the sense that the volume of an object
approaching the $r = M_0/2$ surface is not shrunk to zero, and
therefore the energy density of the cosmic fluid remains finite.
On the other hand, the pressure diverges at $r = M_0/2$ together
with the Ricci scalar $R$ \cite{sus:1985,ml:2006}. So, while the
McVittie metric represents some kind of strongly gravitating
central object, its physical interpretation is not totally clear
and remains the subject of debate \cite{sus:1985, no:1999}. The
only exception is the Schwarzschild- de Sitter (or
K\"{o}ttler)
metric, which corresponds to a special case of the McVittie
metric and represents a genuine black hole of constant mass
embedded in a de Sitter background \cite{sus:1985,
no:1998,fa:2007}.

Because the McVittie solution can not describe a black hole in a
FRW universe (with the exception of the
Schwarzschild-de Sitter
one), we turn our attention to the new solutions of
Ref.~\cite{fa:2007}, described by the generalization of the
McVittie metric \cite{footnote1} \begin{eqnarray}
 \label{eq:DS-dt}
ds^2&=&-\frac{\left[1-\frac{M\left(t\right)}{2a\left(t\right)r}
\right]^2}{\left[1+\frac{M\left(t\right)}{2a
\left(t\right)r}\right]^2} dt^2
  +{a\left(t\right)^2}\left[1
+\frac{M\left(t\right)}{2a\left(t\right)r} \right]^4 \nonumber\\
&&\nonumber \\
  &&\cdot\left(dr^2+r^2d\Omega^2 \right) \;,
 \end{eqnarray} in the background of an imperfect fluid with a
radial heat flux and, possibly, a radial mass flow simulating
accretion onto a black hole embedded in a generic FRW
universe.

In the McVittie metric, the mass coefficient $M$ is constant,
which expresses the assumption (hereafter referred to as the
``McVittie condition'') that $G_0^1=0$, where $G_{\mu\nu}$ is
the
Einstein tensor, and therefore to vanishing radial mass flow
onto the central object, $T_0^1=0$, and to no accretion. $M(t)$
corresponds \cite{fa:2007} to the Hawking-Hayward quasi-local
mass of the metric~(Eq.~\ref{eq:DS-dt}) and, therefore, the McVittie
no-accretion condition corresponds to constant physical mass of
the central object, which is static even when the universe
expands or contracts. Here instead, following \cite{fa:2007},
the black hole is allowed to accrete cosmic fluid and $M(t)$
changes with time.

Two distinct processes concur in the time variation of the black
hole mass: the first is the
radial flow of energy onto the black hole
due to the radial motion of the cosmic fluid toward the horizon
(this is a simplified version of the accretion process
considered by Babichev {\em et al.}), possibly supplemented by a
radial heat flux, which necessarily requires an imperfect fluid
description. The second process is the variation of the mass
enclosed by the black hole apparent horizon due to the motion of
the horizon itself; this process is absent in the McVittie
solution (in which the singular surface $r=M_0/2$ does not
expand) and in the static black hole considered by Babichev {\em
et al.}, but is present for any dynamical horizon even in the
absence of the first accretion process.
 However, unless the total mass of the central object is zero
(which could happen, {\em e.g.}, for a spherical shell of exotic
matter), a radial energy flow is to be expected due to its
gravitational pull. The solutions of \cite{fa:2007} exhibit
accretion onto the central black hole due to radial mass flow
and/or to a radial heat flux.

The components of the Einstein tensor for the
metric~(\ref{eq:DS-dt}) are \cite{fa:2007}
\begin{eqnarray} &&
G_0^0 = -\frac{3A^2}{B^2}\left(
\frac{\dot{a}}{a}+\frac{\dot{m}}{rA} \right)^2 \;, \\
&&\nonumber \\ && G_0^1 = -\frac{2m}{ r^2 a^2 A^5 B}\left(
\frac{\dot{a}}{a}+\frac{\dot{m}}{m} \right) \;, \\ &&\nonumber
\\ && G_1^1 = G_2^2=G_3^3= -\frac{A^2}{B^2}\left\{ 2\,
\frac{d}{dt} \left( \frac{\dot{a}}{a} +\frac{\dot{m}}{rA}
\right)  \right.\nonumber \\ && \nonumber \\ && \left. + \left(
\frac{\dot{a}}{a} +\frac{\dot{m}}{rA} \right)\left[ 3 \left(
\frac{\dot{a}}{a} +\frac{\dot{m}}{rA} \right)
+\frac{2\dot{m}}{rAB} \right]\right\} \;,
\end{eqnarray}
where $ A\equiv 1+\frac{m}{2r}$, $ B \equiv 1 -
\frac{m}{2r}$, and $m(t) \equiv M(t)/a(t)$. Assuming that there is
no mass flow but that there is a radial heat flux in the
stress-energy tensor
\begin{equation}
T_{\mu\nu}=\left( P+\rho \right) u_{\mu}u_{\nu}
+Pg_{\mu\nu}+q_{\mu}u_{\nu} +q_{\nu}u_{\mu} \;,
\end{equation}
where $u^{\mu}=\left( \left|-g_{00}\right|^{-1/2}, 0,0,0
\right)$ and $q^{\mu}=\left(0, q ,0,0 \right)$, the Einstein
equations are
\begin{eqnarray}
G_0^1 &=& -8\pi \frac{A}{B}\, q\;, \\
&&\nonumber \\
G_0^0 &=& -8\pi \, \rho \;, \\
&&\nonumber\\
G_1^1 &=& G_2^2= G_3^3 = 8\pi \, P\;.
\end{eqnarray}
For the constant equation of state $P=w\, \rho$ ($w=$const.),
combined with the $ \left( 0,0 \right)$ and $ \left(1,1 \right)$
Einstein equations to obtain $G_1^1+w\, G_0^0=0$, one derives
\begin{eqnarray}
&& 2 \frac{d}{dt} \left(
\frac{\dot{a}}{a}+\frac{\dot{m}}{rA} \right) + \left(
\frac{\dot{a}}{a}+\frac{\dot{m}}{rA} \right) \left[ 3\left( w+1
\right) \left( \frac{\dot{a}}{a}+\frac{\dot{m}}{rA} \right)
\right. \nonumber \\
&&\nonumber \\
&& \left.
+\frac{2\dot{m}}{rAB} \right]=0
\end{eqnarray}
and
\begin{equation} \label{heatflux}
q=\frac{-\dot{M}}{4\pi r^2 a^3 A^6} \;,
\end{equation}
which expresses the fact that an ingoing flow
($q<0$) corresponds to an increase of the quasi-local mass
$\dot{M}>0$, and vice-versa.

As $r\rightarrow +\infty$, the scale factor of the universe is
given by the familiar expressions

\begin{eqnarray} \label{eq:w-m} a(t) &= & a_0 \left( t_{rip}
-t\right)^{\frac{2}{3\left( w+1 \right) } } \;\;\;\;\;\;\;\;
\mbox{if}\;\;w<-1 \;, \label{eq:bigrip}\\ &&\nonumber \\ a(t) &=
& a_0 \, \mbox{e}^{H_0 t} \;\;\;\;\;\;\;\; \mbox{if}\;\;w=-1
\;,\\ &&\nonumber \\ a(t) &= & a_0 \, t^{ \frac{2}{3 \left( w+1
\right)}}
 \;\;\;\;\;\;\;\; \mbox{if}\;\;w > -1 \;, \label{eq:w-mm}\end{eqnarray} where
$t_{rip}$ denotes the Big Rip. It is easy to see that the previous
form of $a(t)$ with the choice $ M(t)= M_0\, a(t) $ (with $M_0$ a
positive constant) and the corresponding heat flux density
\begin{equation} q=\frac{ -M_0 H }{ 4\pi r^2a^2A^6 } \;,
\end{equation} constitutes a solution of the {\em full} Einstein
equations. The energy density and pressure depend on both the
radius and the comoving time, $\rho=\rho\left( t, r \right)$, $
P=P\left( t, r \right)$, and \begin{equation} \rho \left( t, r
\right)=\frac{3A^2}{B^2}\left(
\frac{\dot{a}}{a}+\frac{\dot{m}}{rA} \right)^2 \end{equation} is
non-negative.

Thanks to the particular form of the metric coefficient $M(t)$,
this solution is conformal to a Schwarzschild black hole with
conformal factor equal to the scale factor of the background
universe $a(t)$ \cite{fa:2007,mybh}, a feature
shared with the
Sultana-Dyer solution \cite{sul:2005} ({\em cf.}
also the
solutions of Ref.~\cite{ml:2006}). The mass
coefficient $M(t)$
coincides with the Hawking-Hayward quasi-local mass of the
spacetime~(Eq.\ref{eq:DS-dt}), which should be identified with the
physical black hole mass \cite{mybh}. The fact that the latter
increases in an expanding universe describes the constant
accretion due to the radial energy flux from the surrounding
fluid. The McVittie condition $G_0^1=0$ equivalent to
$\dot{M}=0$ that forbids accretion onto the central object in
the McVittie metric is clearly not imposed in the solutions of
\cite{fa:2007} that we use here, allowing accretion to occur.

These solutions describe a cosmological black hole embedded in a
FRW background and present advantages over the previous
Sultana-Dyer \cite{sul:2005} and McClure-Dyer \cite{ml:2006}
solutions, in the sense that both the energy density and the
pressure are finite near the black hole horizon, and the energy
density is positive-definite \cite{fa:2007}.

More general solutions with a different form of the mass
coefficient and no conformal Killing horizon are  considered
later in Sec.~IV.

\section{Violation of Cosmic Censorship}

The exact solution describing a black hole embedded in a FRW
universe that we adopt in this section corresponds to the
choice
$M(t)=M_0 \, a(t)$ and $a(t) $ given by Eq.~(\ref{eq:bigrip}),
and it can be written as \begin{eqnarray}
 \label{eq:ds-ph}
ds^2&=&-\frac{\left(1-\frac{M_0}{2r}\right)^2}{
\left(1+\frac{M_0}{2r}\right)^2} \, dt^2
  +{a^2\left(t\right)}\left(1+\frac{M_0}{2r} \right)^4
   \nonumber\\ &&\nonumber \\
  &&\cdot\left(dr^2+r^2d\Omega^2 \right)
 \end{eqnarray} in isotropic coordinates. If $a \equiv 1$, then
$r=M_0/2$ is the event horizon of the Schwarzschild black hole;
however, if $a $ is not constant, $r=M_0/2$ is no longer the
event horizon but merely a 2-sphere enclosed within the AH and a
conformal Killing horizon.

In order to investigate the evolution of the black hole mass and
horizon, it is convenient to rewrite Eq.~(\ref{eq:ds-ph}) using
the areal radius ${\tilde{r}} \equiv
r{\left(1+\frac{M_0}{2r}\right)^2}$. Eq.~(\ref{eq:ds-ph}) then
becomes \begin{eqnarray} \label{eq:ds-2m}
ds^2&=&-\left(1-\frac{2M_0}{\tilde{r}} \right)dt^2 + a^2
\left(1-\frac{2M_0}{\tilde{r}} \right)^{-1}d\tilde{r}^2
\nonumber\\ && \nonumber \\ &&+a^2\tilde{r}^2d\Omega^2 \;.
\end{eqnarray} Further setting $ R=a \tilde{r} $,
Eq.~(\ref{eq:ds-2m})  is rewritten in the
Painlev\'{e}-Gullstrand form as \begin{eqnarray} ds^2&=&-\left[
1-\frac{2M_0a}{R} -\left(1-\frac{2M_0a}{R} \right)^{-1} H^2 R^2
\right] dt^2 \nonumber\\ && \nonumber \\ && + \left(
1-\frac{2M_0 a}{R} \right)^{-1} d R^2 \nonumber\\ &&\nonumber \\
&& -2HR \left(1-\frac{2M_0a}{R} \right)^{-1} dt \, dR +
R^2d\Omega^2 \;. \label{eq:ds-2r} \end{eqnarray}

The coordinate system $ \left( t,R ,\theta,\phi \right)$ is not
orthogonal; the cross-term $dtdR $ can be eliminated by introducing
the new time coordinate $\bar{t}$ defined by \begin{eqnarray}
d\bar{t} = \frac{1}{F\left(t, R\right)} \left[dt+\frac{HR}{
\left(1-\frac{2M_0a}{R} \right)^2 -H^2 R^2 } dR \right] \;,
\end{eqnarray}
 where $ F \left( t(\bar{t},R), R \right) $ is an integrating
factor which always exists and solves the partial differential
equation \begin{eqnarray} \partial_{R}\left( \frac{1}{F} \right)
=\partial_t\left[ \frac{1}{F} \frac{HR}
{{\left(1-\frac{2M_0a}{R}\right)^2-H^2R^2}}\right]
\end{eqnarray} in order to make $d\bar{t}$ an exact
differential. The metric~(\ref{eq:ds-2r}) is cast in the
Nolan gauge form
\begin{eqnarray} ds^2&=&-\left[1-\frac{2M_0a}{R}
-\left(1-\frac{2M_0a}{R}\right)^{-1}H^2 R^2\right] F^2d\bar{t}^2
\nonumber\\ && \nonumber \\ &&+\left[1-\frac{2M_0a}{R}
-\left(1-\frac{2M_0a}{R}\right)^{-1} H^2 R^2\right]^{-1}d R^2
\nonumber\\
&&\nonumber \\
&& + R^2d\Omega^2 \;,
\label{eq:ds-F0}
\end{eqnarray}
where $H(t)$ and $F(t,R)$ are now functions of $\bar{t}$ and $R$.
The Schwarzschild solution in Schwarzschild coordinates is recovered
by setting $a\equiv 1$ (and, consequently, $H=0$ and $R=\tilde{r}$),
while the Schwarzschild-de Sitter solution is recovered by using
Eq.~(1) and Eq.~(13); further, the latter reduces to the de Sitter
solution (in static coordinates) if one also sets $M_0=0$.

In the Schwarzschild-de Sitter spacetime, there exists a black hole
apparent horizon (AH) and a cosmic AH. The metric~(\ref{eq:ds-F0})
describes a spherically symmetric and dynamical black hole in a FRW
background more general than de Sitter. In this case, the event
horizon may not be well-defined, but the AH always exists. The AH is
a marginally trapped surface with vanishing expansion and has been
argued to be a causal horizon for a dynamical spacetime. The AH is
associated with the Hawking temperature, gravitational entropy and
other thermodynamical aspects \cite{ward:2000,bak:99, cai:05,
recent}. The first law of thermodynamics for the AH has been derived
not only in general relativity but also in several other theories of
gravity including the Lovelock, nonlinear, scalar-tensor, and
braneworld theories
\cite{gong:07,ak:07,caic:07,caicao:07,akc:07,she:07,shey:07}. In
view of this point, in order to investigate the evolution of the
black hole mass, we calculate the energy contained inside the AH. To
this end, let us first calculate the radius of the black hole AH:
Eq.~(\ref{eq:ds-F0}) yields
\begin{equation} \label{eq:irs}
1-\frac{2M_0a}{R} \mp HR=0\;,
\end{equation}
 as the equation of the
AH (this expression appears also in Ref.~\cite{DiCriscienzoetal07},
and it can also be derived from Eq.~(\ref{eq:ds-ph}) by replacing
$r$ with $R$). We discard the solution corresponding to the lower
sign and to a negative radius, then Eq.~(\ref{eq:irs}) has the two
physical roots
\begin{eqnarray}
R_{c}&=&{\frac{1}{2H}}{\left(1+\sqrt{1-8M_0\dot{a}} \, \right)}
\;, \\ && \nonumber\\
R_{b}&=&{\frac{1}{2H}}{\left(1-\sqrt{1-8M_0\dot{a}} \, \right)}
\;, \label{Rb}
\end{eqnarray}
where an overdot denotes
differentiation with respect to the comoving time $t$ and the
explicit time-dependence of $H$ and $\dot{a}$ contains the time
evolution of the AH in which we are interested. $R_{c}$ and
$R_{b}$ identify the cosmic AH and the black hole AH,
respectively. When the radius $ R $ crosses the black hole AH
$R_{b}$, the $g_{00}$ and $g_{11}$ terms in the line
element~(\ref{eq:ds-F0}) change sign, revealing that the AH is a
one-way membrane for particles falling into the black hole. A
calculation of the null surface of equation $
g^{\mu\nu}\frac{\partial f}{\partial x^{\mu}}\frac{\partial
f}{\partial x^{\nu}}=0$ shows that the AH is a null 2-sphere.
The coordinate transformations yield the correspondence
\begin{eqnarray}
r=M_0/2 \;\; \Leftrightarrow \;\;
\tilde{r}=2M_0 \;\; \Leftrightarrow \;\; R=2M_0a=2m_H(t)
\end{eqnarray}
where $m_H=M$ is the Hawking-Hayward quasi-local
mass (see the end of this section for a discussion). So, the
conformal Killing horizon $r=M_0/2$ is no longer the black hole
event horizon but is instead a 2-sphere contained inside the AH.
In fact, it follows from Eq.~(\ref{eq:irs}) (with the upper
sign) that the roots of this equation satisfy $R_{c,b}
>2M_0a=2m_H$.
Moreover, the energy density $\rho$ and the pressure $P$ are
both regular on the AH and $\rho$ is positive-definite.

We are interested in the accretion of phantom energy onto the
black hole; the black hole AH and the cosmic AH computed from
Eq.~(\ref{eq:w-m}) are plotted in Fig.~(\ref{fig:R-t}) for the
parameter choice $a_0=1, M_0=1, t_{rip}=1$, and $ w=-2$. As is
clear from this figure, in our solution the black hole AH increases while the
cosmic AH decreases as the universe evolves. If the future
universe is dominated by phantom energy, then there is a time
$t_*$ at which these horizons coincide, which is given by
$  \dot{a}(t_{*})=\frac{1}{8M_0}$, or
\begin{equation}
t_*=t_{rip}-\left(
\frac{16a_0M_0}{3\left| w+1\right|}\right)^{\left| \frac{3\left(
w+1 \right)}{3w+1} \right|} \;. \label{eq:tstar}
\end{equation}
At the comoving time $t_*$, the critical AH radius is $
R_{cr}=\frac{1}{2H}$. At times $t>t_*$ both horizons disappear
and the black hole singularity becomes a naked singularity. We
stress that this happens {\em before} the Big Rip is reached.

The appropriate notion of mass for a cosmological black hole is
the Hawking-Hayward quasi-local mass which, for the metric
(\ref{eq:DS-dt}) reduces to $m_H(t)=M(t)$ and, for the
cosmological black hole~(\ref{eq:ds-ph}), to
\begin{equation}
m_H(t)=M_0 \, a(t) \;.
\label{eq:mH}
\end{equation}
For our solution this is always increasing in an
expanding universe, even when the black
hole AH disappears (the computation of this mass
 does not depend on the presence of an AH), a fact that can be
interpreted as the energy flow continuing to fall onto the naked
singularity after the horizon has disappeared.

The  conclusion that the black hole AH and the
Hawking mass {\em
increase} in a phantom-dominated universe running toward the
Big Rip is the opposite of what Babichev {\em et al.}
\cite{babi:04} find by extrapolating rigorous results obtained
for the accretion of a phantom test fluid onto a
static
Schwarzschild black hole. They find that the Schwarzschild mass
{\em decreases} because the accretion rate $\dot{m} \propto
\left( P+\rho \right)<0$. However, this result can not be generalized
directly to a cosmological, time-dependent, black hole.  The
reason for the discrepancy is that the
Hawking-Hayward
quasi-local mass $m_H=M$ (which implies the subtraction
of a background energy and reduces to the Schwarzschild
mass when $a\equiv 1$) depends only on the energy density $\rho$
but not on the pressure $P$ \cite{Hayward}. Since $\rho>0$, it
does not matter how negative $P$ is, and $\dot{m}_H $ is always
positive. The Schwarzschild mass notion is not
appropriate, nor defined, for
a  black hole in an  asymptotically FRW spacetime.

The quasi-local mass has been proved to be superior to other notions of
mass, which have generated some confusion in the literature, in
the study of the Hawking temperature of cosmological black holes
\cite{mybh}. Nevertheless, it may be interesting to try out a
different notion of mass. An obvious candidate is the
(generalized) Misner-Sharp mass that has been used in black hole
thermodynamics in relation with energy flows through the AH
\cite{gong:07, DiCriscienzoetal07}. While we do not want to fuel
a debate on which is the most convenient notion of mass, and for
which purposes, we present the time evolution of this
generalized Misner-Sharp  mass contained inside the
AH. For a
spherically symmetric space-time with line element $ds^2
=h_{\mu\nu}dx^{\mu}dx^{\nu}+ R^2d\Omega^2$, the generalized
Misner-Sharp mass is $M_{MS}=R\left( 1- h^{\mu\nu}
\,R_{,\mu}
R_{,\nu}\right) /2$ \cite{ms:64}. At the AH, it is $ h^{\mu\nu}
R_{,\mu} R_{,\nu} = 0$ and the generalized Misner-Sharp mass
inside the AH (also computed in Ref.~\cite{DiCriscienzoetal07})
is
simply \begin{equation} \label{eq:MR} M_{MS}={R_{b}}/2
\end{equation} for all times for which the expression~(\ref{Rb})
of $R_b$ is well-defined. Both the quasi-local and the generalized
Misner-Sharp mass of the black hole increase with time: this is at
odds with the decreasing mass property found in Ref.~\cite{babi:04}
(the difference between Misner-Sharp and Hawking-Hayward masses can
be seen also in the Schwarzschild-de Sitter metric).
Eqs.~(\ref{eq:mH}) and (\ref{eq:MR}) show that a discussion of the
time evolution of black hole masses is meaningless until the correct
notion of mass is identified, and the choice is non-trivial for a
time-dependent cosmological black hole.

\begin{figure} \includegraphics[width=6.5cm]{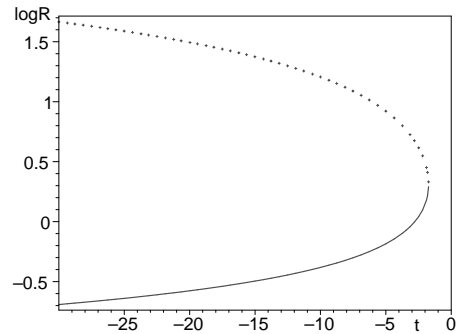}
\caption{The black hole apparent horizon (solid line) increases
while the cosmic apparent horizon (dotted line) decreases with
the accretion of phantom energy onto the black hole. There is a
time at which the two horizons coincide and after which both
horizons disappear and the black hole singularity becomes a
naked singularity. The plots correspond to the equation of state
parameter $w=-2$.} \end{figure}

\section{A more general class of solutions}

One may wonder whether the conclusions reached in the previous
section are due to the peculiar form of the solution employed,
in particular the choice of Hawking-Hayward quasi-local mass
scaling as the scale factor, $M(t)=M_0 a(t)$. The answer is
negative, as we shall see below. The exact solution studied in
the previous section can be generalized to an arbitrary form of
the Hawking mass $m_H=M(t)$ \cite{fa:2007} (in this case, the
2-sphere $r=M/2$ is no longer a Killing horizon). Consider again
Eq.~(\ref{eq:DS-dt}) with a {\em general} form of $M(t)$; by
defining the areal radius $\tilde{r}=r\left(
1+\frac{M(t)}{2ra(t)} \right)^2$ as before and $m(t)\equiv
M(t)/a(t)$, the metric ~(\ref{eq:DS-dt}) is rewritten as
\begin{eqnarray} ds^2&=&-\left[ 1-\frac{2m}{\tilde{r}} -\frac{
a^2 \dot{m}^2 }{ 1-\frac{2m}{\tilde{r}}} \left(
1+\frac{m}{2\tilde{r}} \right)^2 \right] dt^2 \nonumber \\
&&\nonumber \\ &&+ a^2 \left(1-\frac{2m}{\tilde{r}} \right)^{-1}
d\tilde{r}^2 \nonumber\\ && \nonumber \\
&&+a^2\tilde{r}^2d\Omega^2 - \frac{2\dot{m} a^2 \left(
1+\frac{m}{2r} \right)}{1-\frac{2m}{\tilde{r}}} \, dtd\tilde{r}
\;. \label{add2} \end{eqnarray} Introducing the comoving
$R\equiv a\tilde{r}$, in terms of which
$dr=\frac{dR}{a}-H\tilde{r}dt$, Eq.~(\ref{add2})  is turned into
the Painlev\'{e}-Gullstrand form
\begin{eqnarray}
 ds^2 & = & -\left[ 1-\frac{2 M}{R} -\frac{ \left( HR +
\dot{m}a\sqrt{ \frac{ \tilde{r}}{r}} \right)^2 }{ 1-\frac{2M}{R}
} \right] dt^2 \nonumber \\
&&\nonumber \\
& + &
\frac{dR^2}{1-\frac{2M}{R}}+R^2d\Omega^2 \nonumber \\
& - &  \frac{2}{1-\frac{2M}{R}} \left( HR+\dot{m}a\sqrt{
\frac{\tilde{r}}{r}} \, \right) dt \, dR \;.
\end{eqnarray}
Further  setting $ A(t,R) \equiv 1-2M/R $ and $ C(t,R) \equiv
HR +  \dot{m}a\sqrt{ \frac{\tilde{r}}{r} } $ and defining the
time
coordinate $\bar{t}$ by \begin{equation} d\bar{t}=\frac{1}{F}
\left( dt +\frac{C}{A^2-C^2}\, dR \right) \;, \end{equation}
where $F\left( \bar{t},R \right)$ is again an integrating factor
that makes $d\bar{t}$ an exact differential, one has
\begin{eqnarray}
&& ds^2 =- \frac{ A^2-C^2}{A} \left[
F^2d\bar{t}^2 + \frac{ C^2dR^2}{\left( A^2-C^2 \right)^2 }
-\frac{2FB}{A^2-C^2}\,
d\bar{t} \, dR \right] \nonumber \\ &&\nonumber \\ &&
+\frac{dR^2}{A} + R^2d\Omega^2 -\frac{2C}{A} dR \left(
Fd\tilde{t} - \frac{C}{A^2-C^2}\, dR \right)  \;. \end{eqnarray}
The cross terms in $dR \,d\bar{t}$ cancel out and we are left
with the line element in the Nolan gauge
\begin{eqnarray}
&&  ds^2 =- \left( 1-\frac{2M}{R} \right)\left[ 1- \frac{ \left(
HR+\dot{m}a \sqrt{ \frac{\tilde{r}}{r} } \, \right)^2}{\left(
1-\frac{2M}{R} \right)^2 } \right] F^2 d\bar{t}^2 \nonumber \\
&& \nonumber \\
&& +\left( 1-\frac{2M}{R} \right)^{-1} \left[ 1-
 \frac{ \left( HR+\dot{m}a \sqrt{ \frac{\tilde{r}}{r} } \,
\right)^2}{\left( 1-\frac{2M}{R} \right)^2 } \right]^{-1} dR^2
\nonumber \\ &&\nonumber \\ && +R^2 d\Omega^2 \;. \end{eqnarray}
Inspection of this metric locates the horizons at $C=\pm A$, or
\begin{equation} \label{addhorizons} HR+\dot{m}a\sqrt{
\frac{\tilde{r}}{r} } =\pm \left( 1-\frac{2M}{R} \right) \;.
\end{equation} Since the Hawking mass is $M(t)=m(t)a(t)$, the
expression on the left hand side can be written as
 \begin{equation} HR+ M\left( 1+\frac{m}{2r} \right)\left(
\frac{\dot{M}}{M}-H \right) \;, \end{equation} in which the
factor $M\left( 1+\frac{m}{2r} \right)$ measures the deviation
of the radius from $2M$ (with $ r>m/2 \Leftrightarrow R>2M
\Leftrightarrow M\left( 1+\frac{2m}{r} \right) >2M$), while the
factor $ \left( \frac{\dot{M}}{M}-H \right) $ is the difference
between the percent variation of the black hole mass and that of
the scale factor of the cosmic substratum. The vanishing of this
factor can be regarded as the analog of the  condition for
stationary accretion in  a time-dependent background (keeping
with the spirit of the quasi-local mass definition in which a
background quantity is subtracted). Then,
the solution considered in the previous section corresponds to
{\em stationary accretion relative to the FRW background}.

By discarding the
lower sign as usual, Eq.~(\ref{addhorizons}) becomes
\begin{equation}\label{addquadratic} HR^2 +\left[ M\left(
1+\frac{m}{2r} \right)\left( \frac{\dot{M}}{M}-H \right)-1
\right]R +2M=0 \;. \end{equation} Due to the general form of the
function $M(t)$ (not scaling as $ M(t)=M_0 a(t) $), now the
coefficient $\left( 1+\frac{m}{2r} \right)$ appears, in which
$r=r(R)$. Due to this dependence, Eq.~(\ref{addquadratic}) can
not be regarded as a quadratic algebraic equation for the AHs
radii as before. Nevertheless, it is instructive to treat it
formally as such, obtaining the formal roots \begin{eqnarray} &&
R_{c,b}=\frac{1}{2H} \left\{ 1-M\left( 1+\frac{m}{2r}
\right)\frac{\dot{m}}{m} \right. \nonumber \\\ && \nonumber \\
&& \left. \pm \sqrt{ \left[ 1-M\left( 1+\frac{m}{2r}
\right)\frac{\dot{m}}{m} \right]^2-8m\dot{a} } \right\} \;.
\end{eqnarray} Since $r$ depends on $R$, this is only an
implicit equation for the cosmic and black hole AH radii
$R_{c,b}$, but it suffices for our purposes. When the argument
of the square root is positive, there are a cosmic AH at $R_c$
and a black hole AH at $R_b$ while, if this argument becomes
negative because the time derivative $\dot{a}$ increases without
limit as the Big Rip is approached, these AHs coincide at the
common critical radius $ \sqrt{\frac{2M}{H}}$, and then
disappear
leaving behind a naked singularity. This situation occurs, {\em
e.g.}, for the
simple choice of the mass function
\begin{equation}
m(t)=\frac{m_0}{a^{\alpha}(t)} \;\;\;\;\;\; \left( \alpha>1
\right)  \;.
\end{equation}
Then,
$ \left[ 1-M\left(
1+\frac{m}{2r} \right)\frac{\dot{m}}{m} \right]\rightarrow 1$ as
$t\rightarrow t_{rip}^{-}$ and
\begin{equation}
m \, \dot{a}=\frac{
2m_0}{3\left|w+1 \right| a_0^{\alpha-1} } \left( t_{rip}-t
\right)^{\frac{3w+2\alpha+1}{3\left( w+1 \right) }} \;,
\end{equation}
and the heat flux~(\ref{heatflux}) is finite on the AH. If
$\alpha<\frac{3|w|-1}{2}$ (which is compatible with both
requirements $\alpha>1$ and $w<-1$), the horizon
disappears but the central mass does not, leaving a naked
singularity.

\section{Another dynamical phantom black hole}

In order to further clarify the issue, in
this section we construct another  dynamical black
hole solution with phantom scalar field.
To this end, let us first recall the Fonarev solution
\cite {fo:1995}
which is derived from the action
\begin{eqnarray}
\label{eq:action} S=\int
d^4x\sqrt{-g}\left[ R -\frac{1}{2}
\partial_{\mu}\phi\partial^{\mu}\phi-V\left(\phi
\right)\right] \;.
\end{eqnarray}
This action describes an Einstein-scalar system. The potential is
assumed to be exponential
\begin{eqnarray}
\label{eq:potential}
V\left(\phi\right)=V_0e^{-\lambda \phi},
\end{eqnarray}
where $V_0$ and $ \lambda$ are two real constants:
$\lambda$ describes the slope of the potential and
$V_0=V\left( \phi=0 \right)$. Without loss of
generality, we assume that they are both
positive. This potential has been investigated at
length in the context of  homogenous cosmology
\cite{mae:1992}. The Einstein equation and
Klein-Gordon equations are
\begin{eqnarray}
\label{eq:einstein}
G_{\mu\nu}=\partial_{\mu}\phi
\partial_{\nu}\phi-\frac{1}{2}g_{\mu\nu}\partial_{\alpha}
\phi\partial^{\alpha}\phi-g_{\mu\nu}V \;,
\end{eqnarray}
\begin{eqnarray}
\label{eq:phi}
\nabla_{\alpha}\nabla^{\alpha}\phi-\frac{d
V}{d\phi}=0 \;.
\end{eqnarray}
The spherically symmetric solution obtained by
Fonarev is
\begin{eqnarray}
\label{eq:fon}
ds^2&=&a\left(\eta\right)^2\left[-f^2\left(r\right)
d\eta^2+\frac{1}{f^2\left(r\right)}dr^2
+S\left(r\right)^2d\Omega^2\right],\nonumber\\
&&\nonumber \\
\phi&=&\frac{1}{\lambda}\ln
\left[
\frac{V_0\left(\lambda^2-2\right)^2}{2a_0^2
\left(6-\lambda^2\right)} \right] +\lambda\ln
a\nonumber\\&&+\frac{1}{\sqrt{\lambda^2+2}}\ln
\left(1-\frac{2w}{r}\right),
\end{eqnarray}
where
\begin{eqnarray}
f&=&\left(1-\frac{2w}{r}\right)^{\frac{\alpha}{2}},\ \ \ \
\alpha =\frac{\lambda}{\sqrt{\lambda^2+2}}, \nonumber\\
S^2&=&r^2\left(1-\frac{2w}{r}\right)^{1-\alpha}, \ \ \ a=a_0
|\eta|^{\frac{2}{\lambda^2-2}}.
\end{eqnarray}
where $w$ and $ a_0$ are integration constants. The
physical meaning of $w$ will be clear later, while
$a_0$ is a normalization constant for the
scale factor. For simplicity,
hereafter we set $a_0=1$. When $w=0$, the
metric~(\ref{eq:fon})  reduces to the
FRW universe while, when $a=1,
\alpha=1$, it
reduces to the Schwarzschild  solution. So this
solution describes a
black hole embedded in FRW universe.

Now, a {\em dynamical}  phantom black hole solution
can be
constructed from the Fonarev solution via the
transformations
\begin{eqnarray}
\label{eq:trans}
\phi\rightarrow i\phi, \ \ \ \ \lambda\rightarrow
-i\lambda.
\end{eqnarray}
Then, the action becomes
\begin{eqnarray}
\label{eq:actp} S=\int
d^4x\sqrt{-g}\left[ R
+\frac{1}{2}\partial_{\mu}
\phi\partial^{\mu}\phi-V\left(\phi
\right)\right] \;,
\end{eqnarray}
describing an Einstein-phantom system with an
invariant  potential.  The corresponding Einstein and
Klein-Gordon equations are
\begin{eqnarray}
\label{eq:einp}
G_{\mu\nu}=-\partial_{\mu}\phi
\partial_{\nu}\phi+
\frac{1}{2}g_{\mu\nu}\partial_{\alpha}
\phi\partial^{\alpha}\phi-g_{\mu\nu}V,
\end{eqnarray}
\begin{eqnarray}
\label{eq:phip}
\partial_{\alpha}\partial^{\alpha}\phi+\frac{d
V}{d\phi}=0.
\end{eqnarray}
Therefore,  we  construct a dynamical solution
representing a black hole immersed in
a phantom scalar field fluid, and corresponding to

\begin{eqnarray}
\label{eq:fonas}
ds^2&=&a\left(\eta\right)^2
\left[-f\left(r\right)^2d\eta^2
+\frac{1}{f\left(r\right)^2}dr^2
+S\left(r\right)^2d\Omega^2\right],\nonumber\\
&&\nonumber \\
\phi&=&\frac{1}{\lambda}
\ln\frac{V_0\left(\lambda^2
+2\right)^2}{2\left(\lambda^2+6\right)}-\lambda\ln
a\nonumber\\&&-\frac{1}{\sqrt{\lambda^2-2}}\ln
\left(1-\frac{2w}{r}\right) \;,
\end{eqnarray}
where
\begin{eqnarray}
\label{eq:where}
f&=&\left(1-\frac{2w}{r}\right)^{\frac{\alpha}{2}}
\;,\ \ \ \
\alpha=-\frac{\lambda}{\sqrt{\lambda^2-2}} \;,
\nonumber\\
S^2&=&r^2\left(1-\frac{2w}{r}\right)^{1-\alpha} \;, \
\ \
a=\frac{1}{\eta^{\frac{2}{\lambda^2+2}}} \;.
\end{eqnarray}
We assume that $\lambda>\sqrt{2}$. Next, let us
investigate the physical meaning of the constant $w$.
When $\lambda>>\sqrt{2}$, we have $a=1 $ and
$ \alpha=-1$, hence the metric becomes
\begin{eqnarray}
\label{eq:metric}
ds^2&=&-\left(1-\frac{2w}{r}\right)^{-1}
d\eta^2+\left(1-\frac{2w}{r}\right)dr^2
\nonumber\\
&&\nonumber
\\
&&+r^2\left(1-\frac{2w}{r}\right)^{2}d\Omega^2 \;.
\end{eqnarray}
The coordinate transformation
\begin{eqnarray}
\label{eq:coordi}y=r\left(1-\frac{2w}{r}\right)
\end{eqnarray}
transforms Eq.(\ref{eq:metric}) into
\begin{eqnarray}
\label{eq:metricsch}ds^2&=&-\left(1+\frac{2w}{y}\right)d\eta^2+\left(1+\frac{2w}{y}\right)^{-1}dy^2
\nonumber\\&&+y^{2}d\Omega^2 \;,
\end{eqnarray}
which  is the Schwarzschild solution with the
mass $M=-w$. Therefore,  the  parameter $w$
corresponds to the  negative mass.
Hereafter, we use the symbol $-M$ instead of
$w$.

In order to study whether the  singularity can ever
be naked during the
evolution of the phantom field, we
calculate the  black hole AH.
Using the parameters $M$ and $\alpha$,
the phantom black hole solution can be written as
\begin{eqnarray}
\label{eq:M}
ds^2&=&\frac{1}{\eta^{\frac{2\alpha^2-2}{2\alpha^2-1}}}
\left[-\left(1+\frac{2M}{r}\right)
^{\alpha}d\eta^2\right.\nonumber\\
&&\nonumber \\
&&\left.+\left(1+\frac{2M}{r}\right)^{-\alpha}dr^2
\right.\nonumber\\
&&\nonumber \\
&&\left.+r^2\left(
1+\frac{2M}{r}\right)^{1+\alpha}d\Omega^2\right] \;.
\end{eqnarray}
By replacing the conformal time $\eta$ with the
comoving time $t$, we can rewrite the metric as
\begin{eqnarray}
\label{eq:cosmic time}
ds^2&=&-\left(1+\frac{2M}{r}\right)
^{\alpha}dt^2+a\left(t\right)^2\left[
\left(1+\frac{2M}{r}\right)^{-\alpha}dr^2
\right. \nonumber\\
&&\nonumber \\
&&\left.
+r^2\left(1+\frac{2M}{r}\right)^{1
+\alpha}d\Omega^2\right],\nonumber\\
&&\nonumber \\
a\left(t\right)&=& \left(t_0-t\right)^{
-\frac{2\alpha^2-2}{\alpha^2}} \;.
\end{eqnarray}
The scale factor tells us that the constant $t_0$ has
the physical meaning of the moment at which the Big
Rip happens. We have
$\alpha<-1$ because of $\lambda>\sqrt{2}$.
The exponent $\alpha$ is determined by
the slope of the potential.  It is apparent that,
when $M=0$, the metric~(\ref{eq:cosmic time})
describes a phantom-dominated Universe.
For simplicity,  we set $\alpha=-3$, {\em i.e.},
$\lambda={3}/{2}$. Then,
the metric~(\ref{eq:cosmic time}) reduces to
\begin{eqnarray}
\label{eq:Metric}
ds^2&=&-\left(1+\frac{2M}{r}\right)
^{-3}dt^2+a\left(t\right)^2\left[\left(
1+\frac{2M}{r}\right)^{3}dr^2
\right.\nonumber\\
&&\nonumber \\
&&\left.+r^2\left(1+
\frac{2M}{r}\right)^{-2}d\Omega^2\right],\nonumber\\
&&\nonumber \\
a\left(t\right)
&=&\left(t_0-t\right)^{-16/9}.
\end{eqnarray}
Using the physical radius $R=ar( 1+2M/r)^{-1}$
instead of thecoordinate radius $r$, we obtain
\begin{eqnarray}
 \label{eq:AH}
&&1+\frac{8Ma}{R}\left(
1+\sqrt{1+\frac{8Ma}{R}}\right)^{-1}\nonumber\\
&&\nonumber \\
&&-\frac{HR}{32}\left(
1+\sqrt{1+\frac{8Ma}{R}}\right)^5=0 \;,
\end{eqnarray}
as the equation of the AH. Here,
$H\equiv\frac{d\ln
a}{dt}$ is the Hubble parameter.  By setting $x\equiv
1+\sqrt{1+\frac{8Ma}{R}}$, one obtains
\begin{equation}
 \label{eq:ah} aMHx^4-4x^2+12x-8=0 \; .
\end{equation}
We find that Eq.~(\ref{eq:ah}) has only two physical roots which
denote the cosmic AH $R_c$ and the black hole
AH $R_b$, respectively. Using Eq.~(\ref{eq:ah}) and
the definition of $x$, we have plotted the relation
between the comoving
time $t$ and the AHs in Fig.~(2). The latter shows
that the
black hole apparent horizon increases while the cosmic apparent
horizon decreases with the blow-up of the phantom
scalar field. There is a moment at which the two
horizons coincide and after which both
horizons disappear and the black  hole singularity
becomes naked.
\begin{figure} \includegraphics[width=6.5cm]{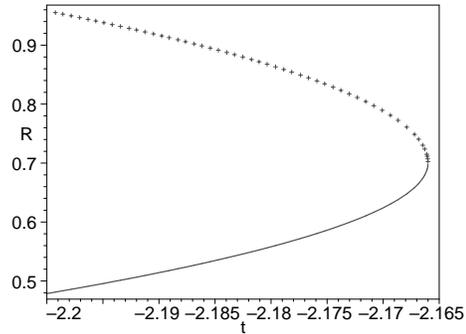}
\caption{The black hole apparent  horizon (solid
line) increases, while the cosmic apparent horizon
(dotted line) decreases with the
divergence of the phantom field. There is a moment at
which
the two horizons coincide and after which
they disappear,  the
black hole singularity becoming naked. These plots
correspond to
$\alpha=-3, t_0=1, M=1$.} \label{fig:R-t}
\end{figure}

\section{Discussion and conclusions}

The well-known McVittie solution can not describe a
black hole
embedded in a FRW universe. On the contrary, the new solutions
of~\cite{fa:2007} do: the surface $r=M_0/2$ is not an event
horizon but a 2-sphere contained inside the black hole AH and a
conformal Killing horizon when $M(t)=M_0a(t)$. The energy
density and pressure are regular at the AH, contrary to the case
of the McVittie metric in which the pressure diverges on the
would-be horizon.  Therefore, we adopt the
solution~(\ref{eq:ds-ph}) of Faraoni and Jacques \cite{fa:2007}
to investigate the problem of phantom matter accretion by black
holes. We take into account the backreaction, which was ignored
in the previous work of Babichev {\em et al.} \cite{babi:04}. In
a low density background this effect can be safely neglected;
however, if the energy density of the background is
large ({\em
e.g.}, if it becomes comparable to the black hole density $
1/M_0^2$), the metric describing this black hole will be
modified significantly. Therefore, backreaction must be part of
the picture.

Using this solution we find that, according to
both the quasi-local and the generalized Misner-Sharp mass
notions, the black hole mass
increases during the cosmic evolution with accretion of the
phantom fluid. First of all, one notes
that more than one notion of mass can be employed, and it is not
trivial to identify the correct one (the Hawking-Hayward
quasi-local mass has proved to be the most convenient in
previous studies \cite{fa:2007, mybh}); this issue is absent
when using a constant mass Schwarzschild black hole as in
\cite{babi:04}.
The discrepancy between our result of increasing
Hawking mass and size of the AH, and the decreasing mass
property found in \cite{babi:04} might be explained by the fact that
the quasi-local mass does not receive contributions from the
negative pressure.

In Faraoni and Jacques \cite{fa:2007},
there is also a solution (eq.~(98)) which yields the
opposite result, {\em i.e.},  the black hole mass
decreases with the accretion of phantom fluid.
However, that solution requires an outgoing radial
energy flux $q=-(P+\rho) u/2$,
which is not  easy to interpret for an accreting
black hole, and could easily explain the decreasing
mass. Here, we have shown that, for an inward energy
flux, the black hole mass increases
even when phantom energy fluid is accreted.

Unfortunately, it is not easy to compare our solution
to the Babichev {\em et al.}\ solution \cite{babi:04}
directly,  as they have
ignored the backreaction effect of expansion, while
we have simplified our model of fluid flow instead.
The comparison is  not straightforward: it is not  a
simple matter of  taking a limit in the relevant
equations, because here the
relevant black hole mass is taken as the
Hawking-Hayward quasi-local mass for the cosmological
black hole immersed in a fluid that sources the
surrounding FRW universe-this is forced upon us
when looking for a meaningful mass concept to use.
When  comparing with \cite{babi:04}, one must replace
this  quasi-local mass with
the  Schwarzschild  mass and neglect its time
dependence, which implies setting $a$ to a constant
and, in turn, means that the background fluid becomes
a test fluid  (it no longer causes the cosmic
expansion).
The crucial difference between our solutions and
that of \cite{babi:04} is that the black hole mass
is no  longer obtained as a ``mass contrast'' with
respect  to the background, but it will include
both  energy density $\rho$ and
pressure $P$ inside the horizon, thus causing a flow
of accreted total energy that is negative (however,
in this regime,  the effect on the Schwarzschild mass
is neglected).
While we do not doubt that the analysis of
\cite{babi:04} is
correct in the test-fluid regime, it is its
extrapolation to a gravitating fluid and to an
expanding  universe that we question, and we can only
propose  some special classes of exact solutions to
further the  analysis. Because these may be very
special, we do not claim that our is the last
word on this  issue - it is an intermediate step
(but a valuable one because exact solutions are
rare), and we auspicate that more exact solutions
incorporating an explicit radial mass flow be found
to discuss this issue further.

The central, and most interesting point of our study, however, is
not the behavior of the black hole mass but rather the process in
which the black hole AH expands while the cosmic AH shrinks. When
the comoving time derivative of the scale factor approaches the
critical value $ 1/(8\pi M_0)$ at the time $t_*$ given by
Eq.~(\ref{eq:tstar}), the two horizons coincide. At
times $t>t_*$  both horizons disappear and the black
hole singularity becomes a
naked singularity; this happens {\em before} the universe reaches
the Big Rip. The shrinking of the cosmic apparent horizon has
serious consequences. The Bekenstein-Hawking formula $ S=A/4 $
\cite{be:75,haw:76} tells us that the horizon entropy $S$ is
proportional to the area $A$ of its cosmic apparent horizon. So, the
entropy of a phantom energy-dominated universe {\em decreases}, in
violation of the second law of thermodynamics. If the black hole
singularity becomes naked in a finite time well before the Big Rip,
the Cosmic Censorship Conjecture is violated.  By combining these
considerations, one could go as far as reaching the conclusion that
nature may not permit the existence of phantom
energy. There is,
however, another consideration to keep in mind: the
solution~(\ref{eq:ds-ph}) (as well as the similar solutions of
\cite{sul:2005, fa:2007}) is meant to describe a black hole that has
existed forever, not one that is created by gravitational collapse.
According to the expression~(\ref{Rb}) of the AH radius on which we
base our considerations, if this black hole is created together with
the universe in a Big Bang singularity, early on the derivative
$\dot{a}$ was arbitrarily large and the central singularity was not
covered by a horizon, either. Hence, the metric~~(\ref{eq:ds-ph})
describes a naked singularity (violating the Cosmic Censorship
Conjecture) which gets dressed by a horizon shortly after the Big
Bang, and evolves as a black hole spacetime until it becomes again a
naked singularity (violating Cosmic Censorship again). This bizarre
behavior leads one to wonder whether this is a physically meaningful
solution. After all, it is well known that naked singularities form
during simulations of gravitational collapse, but they are not
generic. ``Typical'' choices of initial data result in the collapse
to a black hole and naked singularities appear to be the exception
rather than the rule. It could well be that the
solution~(\ref{eq:ds-ph}) describes a non-physical, fine-tuned
situation with bizarre phenomenology. The more general solutions
considered in Sec.~IV exhibit the same phenomenology and this fact
tends to exclude fine-tuning, but no truly definitive statement can
be made at present to this regard due to the paucity of exact
solutions describing the physical situation at hand.

\begin{figure} \includegraphics[width=6.5cm]{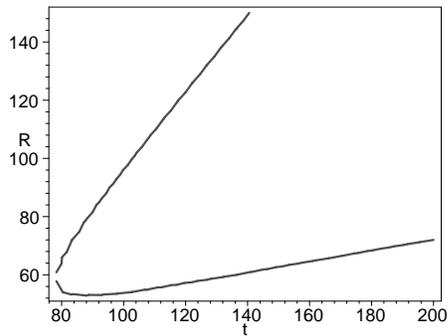}
\caption{The black hole apparent horizon (lower line)
first decreases and then increases, while the cosmic
apparent horizon (upper line) always increases in the
dust-dominated Universe.
There  was a moment at which the two horizons
coincided and before which the singularity was
naked. The plots correspond to $M_0=0.8, a_0=1$.}
\label{fig:R-t-dust}
\end{figure}

Using Eq.~(\ref{eq:w-mm}) and Eq.~(\ref{eq:irs})
with $w>-1$, one finds that the black hole AH first
shrinks and then expands, while
the cosmic AH always expands with the evolution of
the universe, with the cosmic AH expanding faster
than the black hole AH. There
was a moment at which the two horizons coincided and
before which the  singularity was
naked. As an example,  we have plotted in
Fig.~\ref{fig:R-t-dust} the
evolution of the cosmic and black hole AHs in a
dust-dominated universe.

The reason for the black hole AH  shrinking
initially is  that the density, and thus the expansion speed of
the  universe, were very large in the
vicinity of this coincidence. In this sense, the
expansion of the universe ``wins'' over the attraction of the
black hole near this coincidence.

Finally, we comment on the global structure of the cosmological
black holes discussed.  In Figures~4 and~5 we  plot the
Penrose-Carter
diagrams for both the phantom-dominated and the
dust-dominated universe with a black hole,
respectively. The global structure and Penrose diagram  of the
Sultana-Dyer solution has been studied in
\cite{SaidaHaradaMaeda}, and these are similar for  our
conformally Schwarzschild solutions in a dust-dominated
universe (see Fig.~3 of \cite{SaidaHaradaMaeda}).

\begin{figure} \includegraphics[width=6.5cm]{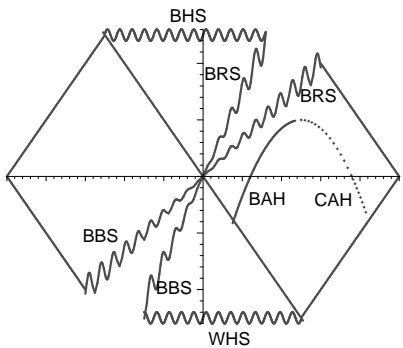}
\caption{The Penrose-Carter diagram for a
phantom-dominated  universe with a
black hole. The diagram is similar to the one for the
Schwarzschild solution except for two sections
removed and two cosmic
singularities added. For brevity, we only label the black hole
singularity (BHS), white hole singularity (WHS),  Big
Bang singularity
(BBS),  and Big Rip singularity (BRS) in the diagram.
The black hole apparent horizon  (BAH) and cosmic apparent
horizon  (CAH) are also shown. } \label{fig:pen-ph}
\end{figure}

\begin{figure} \includegraphics[width=6.5cm]{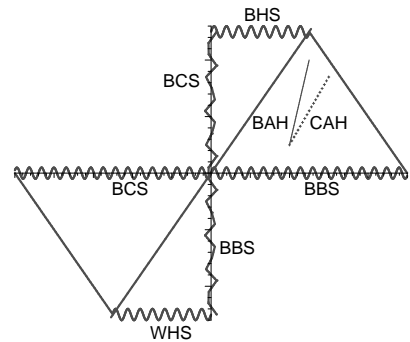}
\caption{The Penrose diagram for a dust-dominated
universe with a black hole. This diagram also
coincides with the one for the
Schwarzschild solution, except for two sections
removed and two cosmic singularities added. We only
label the black
hole singularity (BHS),  white hole singularity
(WHS), Big Bang
singularity (BBS), and Big Crunch singularity (BCS)
in the diagram. The black hole apparent horizon (BAH)
and cosmic apparent horizon
(CAH) are also shown.} \label{fig:pen-mat}
\end{figure}

Hence, the singularity will  never become naked in a
universe
dominated by ordinary (or dark) matter. In a universe driven by
exotic phantom matter, the black hole AH also expands. One then
wonders which mechanism can make the black hole AH to shrink. Recall
that the Schwarzschild anti-de Sitter metric is
\begin{eqnarray} ds^2&=&-\left(1-\frac{2M_0}{r}+
H^2r^2\right)dt^2\nonumber\\&&+\left(1-\frac{2M_0}{r}
+H^2r^2\right)^{-1}dr^2+r^2d\Omega^2 \;.
\end{eqnarray}
In this case, $H$ is a constant and the black hole AH is smaller than
that of the corresponding asymptotically flat black hole.
Accordingly, the generalized Misner-Sharp mass is smaller than
the one of the hole without cosmological constant. We note that
the Schwarzschild-anti-de Sitter spacetime has a remarkable
difference in comparison with Schwarzschild-de Sitter: its
energy density $\rho=-\frac{3}{8\pi}H^2$ is negative. This
solution suggests that if a black hole is surrounded by negative
and decreasing energy density, its mass will decrease
monotonically and will tend to zero in a finite or an
infinite cosmic time. However, even though in the
solutions  considered here the phantom kinetic
energy density is negative, its total energy density
is positive, hence this would not happen.

\begin{acknowledgments}

We thank a referee for insightful
comments leading to improvements in the original
manuscript. Thanks go to R. G. Cai, X. Zhang and H.
S. Zhang  for helpful
discussions. This work is supported by the National Science
Foundation of China under the Distinguished Young Scholar Grant
10525314, the Key Project Grant 10533010, and Grants No. 10575004,
10573027, and 10663001; by the Shanghai Natural Science Foundation
under Grant No. 05ZR14138; by the Chinese Academy of Sciences under
grant KJCX3-SYW-N2; and by the Ministry of Science and Technology
under the National Basic Sciences Program (973) under grant
2007CB815401. V.F. acknowledges the Natural Sciences and Engineering
Research Council of Canada (NSERC).
\end{acknowledgments}

\newcommand\AL[3]{~Astron. Lett.{\bf ~#1}, #2~ (#3)}
\newcommand\AP[3]{~Astropart. Phys.{\bf ~#1}, #2~ (#3)}
\newcommand\AJ[3]{~Astron. J.{\bf ~#1}, #2~(#3)}
\newcommand\APJ[3]{~Astrophys. J.{\bf ~#1}, #2~ (#3)}
\newcommand\APJL[3]{~Astrophys. J. Lett. {\bf ~#1}, L#2~(#3)}
\newcommand\APJS[3]{~Astrophys. J. Suppl. Ser.{\bf ~#1},
#2~(#3)} \newcommand\JCAP[3]{~JCAP. {\bf ~#1}, #2~ (#3)}
\newcommand\LRR[3]{~Living Rev. Relativity. {\bf ~#1}, #2~ (#3)}
\newcommand\MNRAS[3]{~Mon. Not. R. Astron. Soc.{\bf ~#1},
#2~(#3)} \newcommand\MNRASL[3]{~Mon. Not. R. Astron. Soc.{\bf
~#1}, L#2~(#3)} \newcommand\NPB[3]{~Nucl. Phys. B{\bf ~#1},
#2~(#3)} \newcommand\PLB[3]{~Phys. Lett. B{\bf ~#1}, #2~(#3)}
\newcommand\PRL[3]{~Phys. Rev. Lett.{\bf ~#1}, #2~(#3)}
\newcommand\PR[3]{~Phys. Rep.{\bf ~#1}, #2~(#3)}
\newcommand\PRD[3]{~Phys. Rev. D{\bf ~#1}, #2~(#3)}
\newcommand\RMP[3]{~Rev. Mod. Phys.{\bf ~#1}, #2~(#3)}
\newcommand\SJNP[3]{~Sov. J. Nucl. Phys.{\bf ~#1}, #2~(#3)}
\newcommand\ZPC[3]{~Z. Phys. C{\bf ~#1}, #2~(#3)}


\begin{thebibliography}{99}
\bibitem{footnote1} This kind of solution was also hypothesized,
but not pursued, in Ref.~\cite{DiCriscienzoetal07}.
\end{thebibliography}
\end{document}